\documentclass[12pt]{article}
\usepackage{amssymb}
\usepackage{epsfig}
\usepackage{graphicx}
\usepackage{amssymb}
\usepackage{wasysym}
\newcommand{\E}{\mbox{1\hspace{-2.5pt}l}}
\title{Self-assembled granular walkers}
\author{Z. S. Khan${}^1$, A. Steinberger${}^{1,2}$, R. Seemann${}^{1,3*}$ and S. Herminghaus${}^{1}$}
\date{\today}
\begin{document}
\maketitle
\thispagestyle{empty}
\noindent
$^{1}$ Max Planck Institute for Dynamics and Self-Organization, Bunsenstr. 10, 37073 G\"ottingen, Germany.
\\
$^{2}$ Laboratoire de Physique, Ecole Normale Sup\'erieure de Lyon, CNRS UMR~5672, Universit\'e de Lyon, 46 All\'ee d'Italie, F-69364 Lyon, France.
\\
$^{3}$ Experimental Physics, Saarland University, D-66041 Saarbr\"ucken, Germany.
\\
$^{*}$ e-mail: r.seemann@physik.uni-saarland.de
\\
\newpage
\begin{abstract}
\noindent Mechanisms of locomotion in microscopic systems are of
great interest not only for technological applications, but also for
the sake of understanding, and potentially harnessing, processes far
from thermal equilibrium. Down-scaling is a particular
challenge, and has led to a number of interesting concepts including
thermal ratchet systems and asymmetric swimmers. Here we present a
system which is particularly intriguing, as it is
self-assembling and uses a robust  mechanism which can be
implemented in various settings. It consists of small spheres of
different size which adhere to each other, and are subject to an
oscillating (zero average) external force field. An inherent
nonlinearity in the mutual force network leads to force
rectification and hence to locomotion. We present a model that
accounts for the observed behaviour and demonstrates the wide 
applicability and potential scalability of the concept. \\
\end{abstract}

The fascination with small-scale locomotion systems in biology, such
as ciliates, flagellates, and molecular motors, has spurred a large
number of attempts to construct artificial devices with similar
merits \cite{HuangAPL,DreyfusNature,BernaNatMat,BrowneNatNano,YuPhysFlu,QianPRL,
TiernoPRL,ToonderLoC,ZerroukiNature,GhoshNano,ZhangAPL}.
Systems of particles in fluid dynamics settings can exhibit directed motion under 
periodic and symmetric forcing via the symmetry-breaking of surface or streaming flows \cite{AransonSwimmingsnakes, WrightDimer}. Additionally, the controlled transport of particles and structures can be used for applications such as targeted delivery and stirring
in lab-on-a-chip devices \cite{AransonSwimmingsnakes,ChangNatMat,AransonStaticsnakes}.
 Dry frictional ratchet devices composed of aspherical particles with complex interactions have also been shown to convert periodic and symmetric external forces into net locomotion \cite{KudrolliDimer,KudrolliRod}. Here we report the discovery of a strikingly simple ratchet-like system which self-assembles in proper environments. \\

\begin{figure}[p!]
\begin{center}
\includegraphics[width=5in]{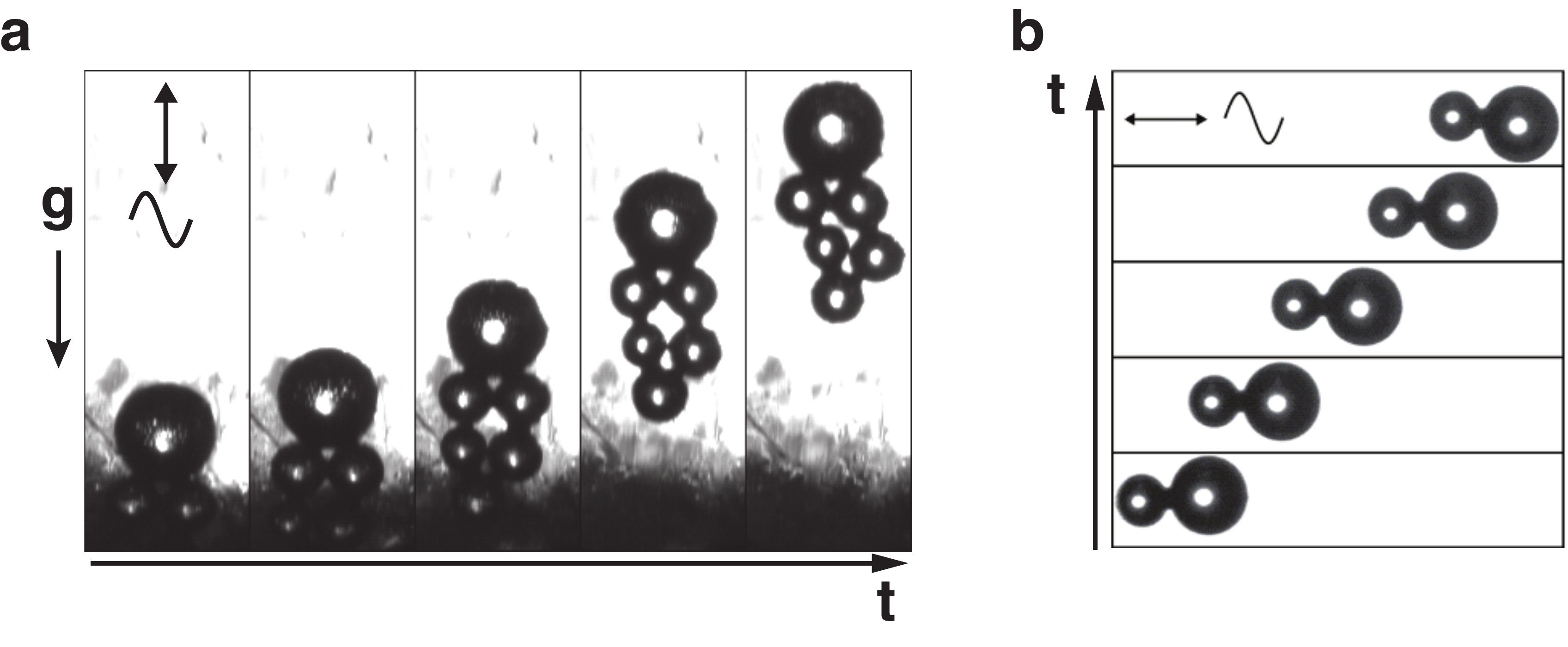}
\caption{{\bf Time series of ratcheting structures.} {\bf a,} This structure was self-assembled from a mixture of glass beads with radii in the size ranges of 0.3-0.315 mm and  0.5-0.59 mm and is migrating up the sidewall of a rectangular polystyrene container under vertical sinusoidal vibration (see supplementary movie M1). The sample was wetted 1\% by volume with a glycerol-water mixture and the container was shaken at a frequency of 170 Hz and a peak acceleration of 16.2 g, where 1 g = 9.8 m/s$^2$. The time interval between images is 2 s. {\bf b,} This artificially assembled dimer, composed of precision spheres with 0.2 mm and 0.3 mm  radii, is migrating along the axis of vibration of a horizontally aligned glass microscope slide which was shaken with a frequency of 80 Hz and a peak acceleration of 4 g (see supplementary movie M2). The substrate was wetted with of silicone oil Si AK5. The time interval between images is 0.3 s.}
\label{Ffig1}
\end{center}
\end{figure}

When a bidisperse mixture of glass beads is moistened by a fluid and
shaken  vertically and sinusoidally, small clusters of beads
occasionally take off from the surface of the pile and rapidly climb
up the container walls {\it against gravity}. These clusters are
held together and against the wall by capillary bridges; they are
led by a large bead with one or more small beads trailing below. In this system, 
the self-assembly of these structures is assisted by the Brazil Nut
Effect \cite{RosatoPRL}, as the large beads are transported to the
top of the pile under vibration. Many different ascending clusters
have been observed, which differed greatly in the number of involved
spheres. Figure \ref{Ffig1}a and Supplementary Movie M1 show a realization of
such an assembly which spontaneously formed and climbed out of a
pile of glass spheres wetted with a glycerol-water mixture. The
upper surface of the granular pile can be seen as the dark region at
the bottom of the images. This effect is robust, as it has been
successfully reproduced using numerous different wetting liquids
(silicone oil, glycerol-water mixtures, ethylene glycol), container
materials (glass and polystyrene) and geometries
(cylindrical, rectangular).\\

In order to investigate this locomotion mechanism in a controlled
setting, we have reproduced this effect with artificially assembled
clusters of precision spheres on a silicone oil wetted, chromium
coated glass substrate. The chromium coating ensured the absence of 
static charging effects on the moving clusters. The substrate was adjusted horizontally, so 
that the effect of gravity on the cluster locomotion can be
neglected. A horizontal harmonic vibration was applied to the
substrate; therefore, the horizontal acceleration of the substrate in the reference frame of the laboratory is $a(t)= a_0 \cos(2\pi\!ft)$, where $a_0$ is the peak acceleration of the substrate and $f$ is the shaking frequency. We
focussed our study on the simplest structure: an asymmetric dimer
built out of one large and one small bead, which we call a walker.
Quite remarkably, the walkers were found to align with the axis of
vibration of the substrate as soon as the vibration was applied, and
to migrate in the direction of the larger sphere. The walkers
travelled with a constant speed when viewed stroboscopically as described in the Methods section; one
example of this motion is shown in figure \ref{Ffig1}b (see Supplementary Movie M2).
\\

Figure \ref{Ffig2} shows our experimental measurements of the
walkers' velocity for vibration frequencies $f$ from 60 to 90 Hz
(symbols) against the peak acceleration $a_0$ of the horizontal
vibration, for a walker with $R_1 = 0.3$ mm and $R_2 = 0.2$ mm. The 
values of the walkers' locomotion velocities were obtained by
averaging a minimum of six measurements, where half of the
measurements were done with the large sphere facing to the left and
half where the large sphere was facing to the right. Both locomotion
directions along the axis of vibration were thereby sampled, thus
ruling out the possible influence of any residual asymmetry of the
setup. The size of our experimental errors corresponds to the
difference between the largest and smallest velocity measurements; 
the experimental uncertainties represent the dispersion of the
velocities due to the presence of dirt or surface irregularities. We
observed a nonlinear dependence of the walkers' locomotion velocity on the peak
acceleration; it rises sharply as the acceleration increases until
it reaches a plateau near $a_0 \approx 6 $ g (or 58.8 m/s$^2$).
Additionally, the data exhibit no significant frequency dependence,
which suggests that the ratio of the applied force to the cohesive
force is a relevant control parameter (as opposed to, for example,
an energy ratio that would imply a dependence on the peak velocity of the substrate
\cite{Fingerle08}). For the following data analysis, we thus
introduce the dimensionless peak force $\Gamma_0$ as the ratio between the peak inertial force and the average capillary
force between the beads and the substrate. For a perfectly wetting liquid such as silicone oil,
\begin{equation}
\Gamma_0 = \frac{a_0 M}{4 \pi \gamma \bar{R}}
\label{E_gam}
\end{equation}
where $M$ is the sum of the masses of the spheres, $\bar{R}$
is the arithmetic mean of their radii, and $\gamma$ is the surface tension
of the wetting fluid \cite{McFarlane}. The inset shows
that the minimum dimensionless peak force required for locomotion to
occur, $\Gamma_{\mbox{c}}$, is also independent of the
vibration frequency within the experimental uncertainties. We
interpret this threshold as being due to finite dry solid friction at the points of contact.\\

We have also observed that the velocity of the walkers increases with the
degree of asymmetry of the dimer. The asymmetry of the structure is
characterised by the angle $\delta$ between the centers of the
spheres and the horizontal as shown schematically in the inset of
figure \ref{Ffig3}; the asymmetry parameter is defined as $\sin\delta = \frac{R_1-R_2}{R_1+R_2}$. 
The velocity of the walkers as a function of $\sin{\delta}$, measured with a fixed 
dimensionless peak force $\Gamma_0 = 0.5$, is
shown in the main panel of figure \ref{Ffig3}. The velocity was
found to vanish for symmetric walkers composed of spheres with identical radii ($\delta = 0$) on a well-levelled substrate. We additionally observed that the symmetric dimers aligned perpendicularly with the external acceleration, in contrast to
the asymmetric walkers which self-oriented in the shaking direction. These observations establish the asymmetry parameter $\sin{\delta}$ as another control parameter.\\

Having identified the relevant parameters, we will now propose a
simple model to explain our experimental observations. The purpose
of our analysis is to show that even for the simplest assumptions 
to be made on the contact characteristics,  including linearity, the
asymmetry of the system inevitably leads to a net motion in one direction.\\

We consider a walker that is aligned with the shaking direction on a horizontal substrate. Assuming that the bead-bead and bead-substrate distances remain fixed, writing down the equations of motion for each bead leads to two sets of equations. 
The first one relates the magnitude of the normal contact forces acting at the three contact points, represented by the 3-component vector $\mathbf{F}:=(F_{1},F_{2},F_{3})$, to their values at rest, represented by the vector $\mathbf{F^0}$, and to the friction forces at the three contact points $\mathbf{F}^{\|}:=(F_{1}^{\|},F_{2}^{\|},F_{3}^{\|})$. Here the indices $1$ and $2$ refer to the contacts between spheres $1$ and $2$ and the substrate, and the index $3$ refers to the contact between the spheres. We obtain for the normal contact forces:
\begin{equation}
\mathbf{F}  = \mathbf{F^0} + \mathbb{D}\mathbf{F}^{\|},
\label{F_eq}
\end{equation}
where $\mathbb{D}$ is a 3x3 matrix which only depends on the angle $\delta$ and on the relative mass difference $\mu = \frac{m_1-m_2}{M}$, where $m_i$ is the mass of bead $i$ \cite{SuppMat}.\\

 The second set of equations describes the rate of change of the sliding velocities at the contact points, $v_i = R_i\dot{\varphi_i} + v_x$ for
$i\in\{1,2\}$ and $v_3 = R_1\dot{\varphi_1} + R_2\dot{\varphi_2}$, where $\dot{\varphi_i}$ is the angular velocity of bead $i$ and $v_x$ is the translational velocity of the walker on the substrate:
\begin{equation}
M\frac{d}{dt}
\left(
\begin{array}{c}
v_1 \\ v_2 \\ v_3
\end{array}
\right) = -Ma(t)\left(
\begin{array}{c}
1  \\ 1 \\ 0
\end{array}
\right) - \mathbb{G}\mathbf{F}^{\|}.
\label{v_diff_eq}
\end{equation}
where $\mathbb{G}$ is a 3x3 matrix which depends upon the moments of inertia of the spheres \cite{SuppMat}.\\ 

For real contacts, the friction force $F_{i}^{\|}$ usually depends on the normal contact force $F_{i}$ and is also an odd function of the sliding velocity $v_i$ \cite{Olsson,Ludema,Mo09}:
\begin{equation}
F_{i}^{\|}(F_{i},-v_i)=-F_{i}^{\|}(F_{i},v_i).
\label{friction}
\end{equation} 
\noindent This type of friction law, associated with the relationship between the normal contact force $F_{i}$ and the three friction forces described by equation \ref{F_eq}, leads to a nonlinearity in the velocities $v_i$ in equation \ref{v_diff_eq}. It is this nonlinearity which allows for mechanical rectification effects, and thus locomotion under a periodic and symmetric inertial force $-Ma(t)$. This is a generic feature that arises even for the simplest friction law obeying relationship \ref{friction}, a friction force with a linear dependence on both the sliding velocity and the normal contact force. Indeed, if we assume a friction law $F_{i}^{\|}=\kappa v_i F_{i}$ for all three contact points, equation \ref{F_eq} leads to a nonlinear expression for the friction forces with respect to the sliding velocities:
\begin{equation}
\mathbf{F}^{\|}=
\mathbb{W}(\E - \mathbb{D}\mathbb{W})^{-1}\mathbf{F^0}
\label{friction_law}
\end{equation}
where the 3x3 matrix
$\mathbb{W}$ has $\kappa v_1$, $\kappa v_2$, and $\kappa v_3$ in its diagonal as
the only non-zero entries.\\

We analytically solved equation \ref{v_diff_eq} with the friction forces given by equation \ref{friction_law} for a quasi-static symmetric excitation $\pm \Gamma_0/\sqrt{2}$, neglecting the weight of the beads in comparison to the capillary forces, so that at rest the normal contact forces $\mathbf{F^0}$ were equal in magnitude and opposed in direction to the capillary forces \cite{SuppMat}. We obtained the following expression for the average velocity of the walker $\bar{v}_x$, averaged over one symmetric excitation, as a function of the squared peak dimensionless force $\Gamma^2_0$ and the asymmetry $\delta$:
\begin{equation}
\bar{v}_x\left(\Gamma^2_0,\delta\right) = \frac{\Gamma^2_0 \sin\delta}{4\kappa} (h_1(\Gamma^2_0,\delta) + h_2(\Gamma^2_0,\delta))
\label{Velocity}
\end{equation}
where
\[h_1(\Gamma^2_0,\delta) = 
\frac{\cos^3\delta (1+\mu\sin\delta)}
{[\cos^2\delta(1+\sin^2\delta)-
\frac{\Gamma^2_0}{8}(1+\mu\sin\delta)^2]^2-[2\cos^2\delta\sin\delta]^2}
\]
and

\[h_2(\Gamma^2_0,\delta) = \frac{4\cos\delta (1+\sin^2\delta)}
{[\cos^6\delta-\frac{\Gamma^2_0}{2}(\mu+\sin\delta)^2][1+3\sin^2\delta]}.
\]
Note that if the beads were to exchange their positions, the average velocity $\bar{v}_x$ changes sign, since $\bar{v}_x(\pi - \delta)=-\bar{v}_x(\delta)$, as was observed experimentally.\\

The average velocity $\bar{v}_x$ (eqn. \ref{Velocity}) is plotted along with the experimental data in figures \ref{Ffig2} and \ref{Ffig3}. In order to fit this expression to our data, the walker asymmetry parameter $\sin{\delta}$ was held fixed at 0.2 in figure \ref{Ffig2}, and the dimensionless force $\Gamma_0$ was held fixed at 0.5 in figure \ref{Ffig3}, in correspondence with the experimental conditions. The fit parameter $\kappa = 0.033$ s$/$mm was used in both figures. 
While we have obtained a single well-defined value for the velocity dependent friction 
coefficient $\kappa$, we cannot compare this finding with literature values because of the unconventional friction law that was implemented to demonstrate the robustness of the rectification 
effect \cite{Ludema}. In figure \ref{Ffig3} we observe that the predicted velocity dependence of the walker on the asymmetry parameter, $\sin{\delta}$, is consistent with the experimentally observed trend. In particular, the locomotion velocity vanishes as the asymmetry parameter goes to zero.
In figure~\ref{Ffig2}, the initial increase of the locomotion
velocity as a function of the dimensionless force $\Gamma_0$ is successfully
reproduced by the theory. However, the threshold for motion at low dimensionless force is not reproduced; the implementation of a more realistic friction law including static friction would be necessary to also predict the onset of motion. Another deviation is observed at higher dimensionless force, as the theoretical expression continues to increase while the experimental values level off and form a plateau around $\Gamma_0=0.6$. 
This may be qualitatively understood as follows.
The model we proposed is only valid as long as the bead-bead and bead-substrate distances remain fixed. As soon as the driving becomes strong enough
to pull the beads off the substrate or from each other, the model doesn't hold anymore, and another mode of motion not described by this model can be triggered. The plateau in the experimental velocities observed at $\Gamma_0 \geq 0.6$  suggests that directed locomotion can also be obtained for other modes of operation than the one we described in the model.\\

\begin{figure}[p!]
\begin{center}
\includegraphics[width=5in]{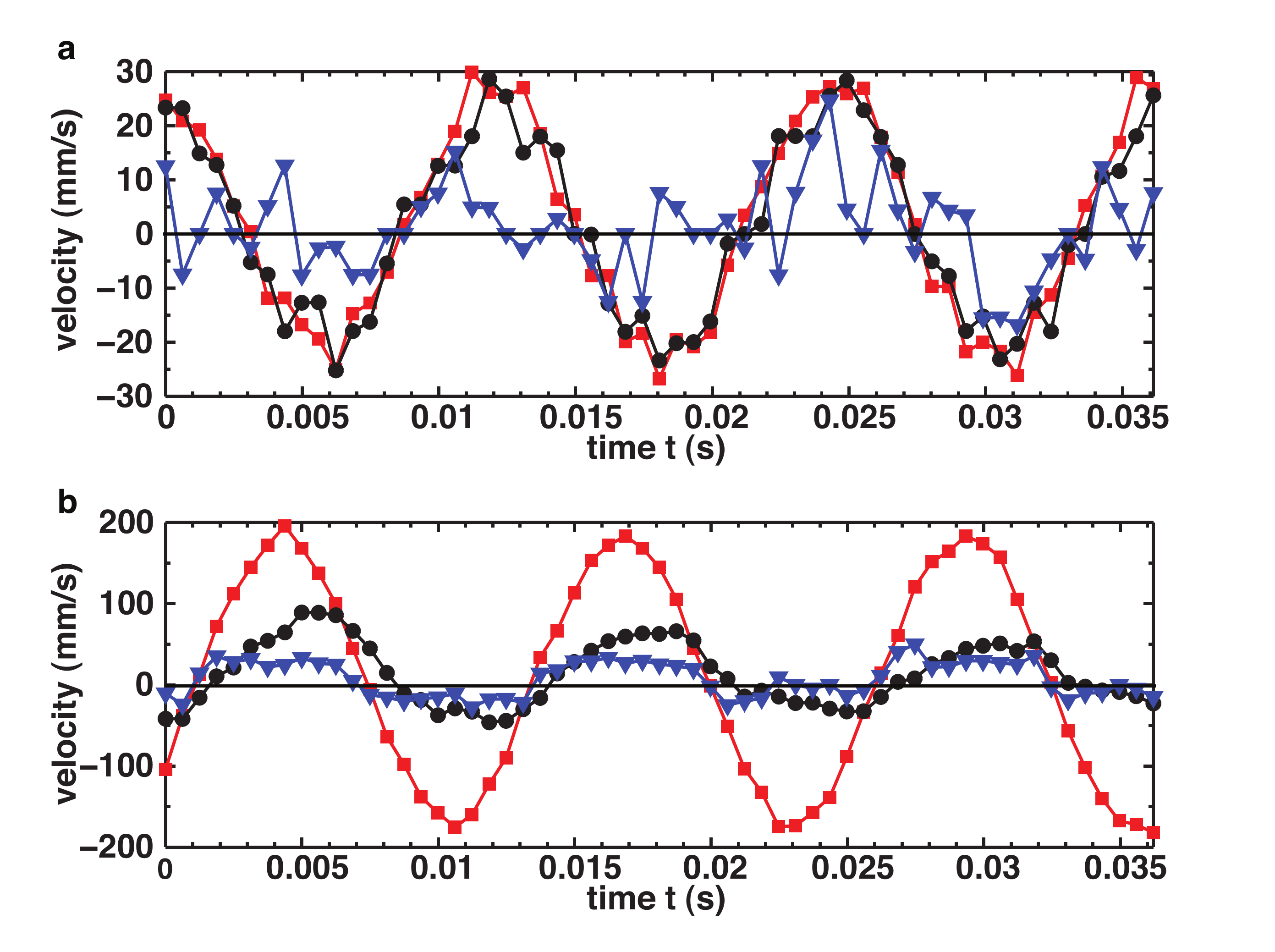}
\caption{{\bf Modes of motion.} The velocity (red squares) of a walker structure
in the reference  frame of the substrate, where $R_1 = 0.6$ mm and
$R_2 = 0.3$ mm. The walker was shaken with a frequency of 80 Hz and:
{\bf a,} a peak substrate acceleration of 2 g, $\Gamma_0 = 0.74$,
using Si AK 5, {\bf b,} a peak substrate acceleration of 10 g,
$\Gamma_0 = 3.5$, using Si AK 10 (see supplementary movie M3).
The walker's velocity (red squares) is displayed alongside the rotation velocity $-R\dot{\varphi}$ of
both the large (black circles) and small
spheres (blue triangles) comprising the walker. The angular
displacements were obtained by tracking the displacements of 0.04 mm
radius glass spheres placed on each bead in the structure and were
recorded  at 1600 fps or 20 images$/$shake. These experiments were performed on the
surface of a glass slide. The mean velocity of the walker
is: {\bf a,} 2.3 mm/s, and {\bf b,} 8.0 mm/s.} \label{Ffig4}
\end{center}
\end{figure}

To investigate this in more detail, we have directly measured the
rotation velocity for each sphere in a walker structure as shown in
figure \ref{Ffig4}. This was done by tracking the angular
displacements of very small tracer particles placed on each of
the two spheres. While obtaining data using this method was very difficult 
for small dimensionless force, good quality data were successfully obtained 
for $\Gamma_0 > 0.7$, for a walker composed of larger spheres.
Figure \ref{Ffig4}a shows that the velocity of
the walker (red squares) is dominated by the rotation of the larger sphere 1
(black circles) at an applied acceleration of 2 g, $\Gamma_0 =  0.74$, and a vibration frequency of
80 Hz. The smaller sphere 2 (blue triangles) shows some sliding
motion as  the difference between the total velocity of the structure
and the rotation velocity gives the sliding velocity. From this data we
conclude that the small sphere is dragged by the larger sphere
during each stroke of the shaker while the large sphere rotates
almost exclusively, keeping firm contact with the substrate. In
contrast, figure \ref{Ffig4}b shows the walker velocity and rotation
velocities of the large and small sphere for an applied acceleration of
10 g, $\Gamma_0 = 3.5$, using a higher viscosity silicone oil, Si AK
10 (see Supplementary Movie M3).  Figure \ref{Ffig4}b shows that there is considerable sliding
motion present in the locomotion of both the large and small spheres,
however the mean motion of the walker is again dominated by rotation
of the larger sphere 1. This was determined by observing that the  mean rotation
velocity is larger (13.3 mm$/$s) than the mean sliding velocity (-5.3 mm$/$s). These measurements
demonstrate that the walkers display different modes of locomotion.\\

Finally, another interesting feature is that the walkers are observed to remain aligned with the direction of the excitation. In order to understand the stability of the walker's direction of locomotion, we have to consider the sideways motion of the walker in the plane of the substrate. Let $\psi$ be the angle between the symmetry axis of the walker, projected onto the plane of the substrate, and the direction of the applied oscillation. If the pair of beads roll sideways without sliding, and the walker's acceleration associated with locomotion is small relative to the acceleration of the substrate, the dynamics of the angle $\psi$ is equivalent to the motion of a parametric pendulum. For small angles $\psi$ and a harmonic driving force oscillating with a frequency $f$, the equation of motion for sideways rolling reduces to a differential equation of Mathieu type \cite{Lachlan}. Defining the dimensionless time variable $\vartheta = \pi\!ft$, this equation can be written in canonical form as 
\begin{equation}
\frac{d^2\psi}{d\vartheta^2} + [U - 4V \cos(2\vartheta)] \psi = 0
\label{Mathieu}
\end{equation}
where $U$ is proportional to the time-averaged driving force, so here $U = 0$. Our system thus dwells on the abscissa of the stability diagram in the $(V,U)$ plane, whose central interval 
$[-0.45;+0.45]$ lies within a stable region \cite{Lachlan}. It can be shown that $|V|<0.4$ for all of the parameters investigated in this work \cite{SuppMat}. Since 
this is well within the stable region, this explains why the walkers remain 
aligned with the direction of the excitation.\\

In conclusion, we have reported the discovery of a novel ratchet system
which is notably robust, self-assembling, self-adjusting to
external driving, functions under a wide range of experimental
conditions, and displays a rich dynamical behaviour with several modes of motion. 
We have presented a basic model of this system that
successfully reproduces the main experimental features for moderate
driving $\Gamma_0$. This model considers only the geometry of the
system, a friction law depending both on the sliding velocity and the normal contact force, 
and the ratio between the driving and
adhesion forces, but doesn't rely on the exact nature of those
forces nor on the average bead size. This suggests that similar
effects could be expected for a wide variety of analogous systems,
with different driving methods (inertial, acoustic, viscous,
electromagnetic), and possibly at a much smaller scale with adhesion arising from van der Waals forces
instead of capillary forces.\\

\noindent{\bf Methods:}\\%

The sample shown in figure \ref{Ffig1}, left panel, consists of a wetted 50:50 mixture of glass spheres with radii ranging from 0.3-0.315 mm and  0.5-0.59 mm, density 2.5 g/cm$^3$, purchased from Whitehouse Scientific. The sample was wetted 1\% by total volume with Glycerol 87\% from Merck,  with density $\rho = 1.23$ g/cm$^3$, surface tension $\gamma = 61 \pm 1$ mN/m and viscosity $\nu =( 9.8 \pm 0.8)\times 10^{-5}$ m$^2$/s. The sample was vertically vibrated in a polystyrene container with square cross-section (10.6 mm long and wide, 44.8 mm high) purchased from VWR International GmbH. \\

The remaining experiments were performed using precision ruby spheres with 0.2, 0.25, 0.3, 0.35, 0.4 and 0.6 mm radii and density 4.0 g/cm$^3$ purchased from Sandoz Fils SA. The experiments shown in figure \ref{Ffig4} used tracer spheres with radii in the range of 0.038-0.045 mm purchased from Whitehouse Scientific. The wetting liquids used were filtered Wacker silicone oils Si AK 5 or Si AK 10. Si AK 5 has a kinematic viscosity $\nu = 5$ mm$^2$/s, surface tension $\gamma =  19.2$ mN/m and density $\rho  = 0.92$ g/cm$^3$. Si AK 10 has $\nu = 10$ mm$^2$/s, $\rho = 0.93$ g/cm$^3$, and $\gamma = 20.2$ mN/m. \\

Vibration was applied using a LDS model V406 electromagnetic shaker. We monitored the applied acceleration using a Kistler triple-axis accelerometer 8690C50 and ensured that the unwanted acceleration in the vertical direction was below 3\% of the driving acceleration.
The horizontal vibration experiments used either Marienfeld glass microscope slides as the substrate (76 mm long, 26 mm wide, 1 mm thick) or a 90 degree glass prism (with a length and height of 25 mm, 35.4 mm hypotenuse). A 50 mm by 25 mm area of the glass slides and one entire face (25 mm by 25 mm) of the glass prism were coated with a 4-5 nm thick layer of chromium using a BOC Edwards Auto 306 Evaporation System  to eliminate static charge effects on the moving clusters of beads. The substrates were wetted with a 4-6 $\mu$m thick layer of Silicone oil.\\

Video imaging was performed using a PCO 1200hs  CMOS camera. PCO Camware software was used for camera control and  image acquisition. Top view experiments (figures \ref{Ffig1}b, \ref{Ffig2} and \ref{Ffig3}) were imaged by mounting the camera onto a Zeiss Stemi 2000 C stereo microscope with a 0.63$\times$ front lens; the time interval between successive pictures was set to an integer number of vibrations, typically one image per shake or one image every 10 shakes, so that the substrate appeared immobile in the time series of pictures. Side view experiments used the camera with an Edmund Industrial Optics 4.5$\times$ zoom lens (figure \ref{Ffig1}a) or an Edmund Industrial Optics 10$\times$ zoom lens (figure \ref{Ffig4}); the frame rate was set to 20 images per shake. The motion of the substrate (figure \ref{Ffig4}) was determined by imaging a ruler fixed to the substrate simultaneously with the walker; from this the phase of the vibration was determined while the vibration amplitude was calculated from measurements of the peak applied acceleration.\\

In order to obtain velocity measurements from the recorded images, the time series of images were opened with ImageJ software (freely available from http://rsb.info.nih.gov/ij/ ). A single threshold was applied to the images to isolate the structure from the background and the particle analysis package was used to determine the centroids of the structures in each image. Further analysis of the time series of structure positions was done using Matlab software.\\


\noindent{\bf Acknowledgements:}\\

The authors thank Udo Krafft for technical assistance,  Elisabeth Charlaix and Mario Scheel for
insightful discussions, and Jonathan Bodenschatz and Michael Wiedecke for help with the setup. 
Z. S. K. acknowledges a fellowship from the Alexander von Humboldt foundation.\\

\noindent{\bf Author Contributions:}\\

Z. S. K., A. S., and R. S. designed experiments; Z. S. K. and A. S. carried out
experiments and analysed data; S. H. and A. S. developed the theory; Z. S. K., A. S., R. S. and
S. H. wrote the paper.\\

%

\noindent
%

%
\begin{figure}[p!]
\begin{center}
\includegraphics[width=5in]{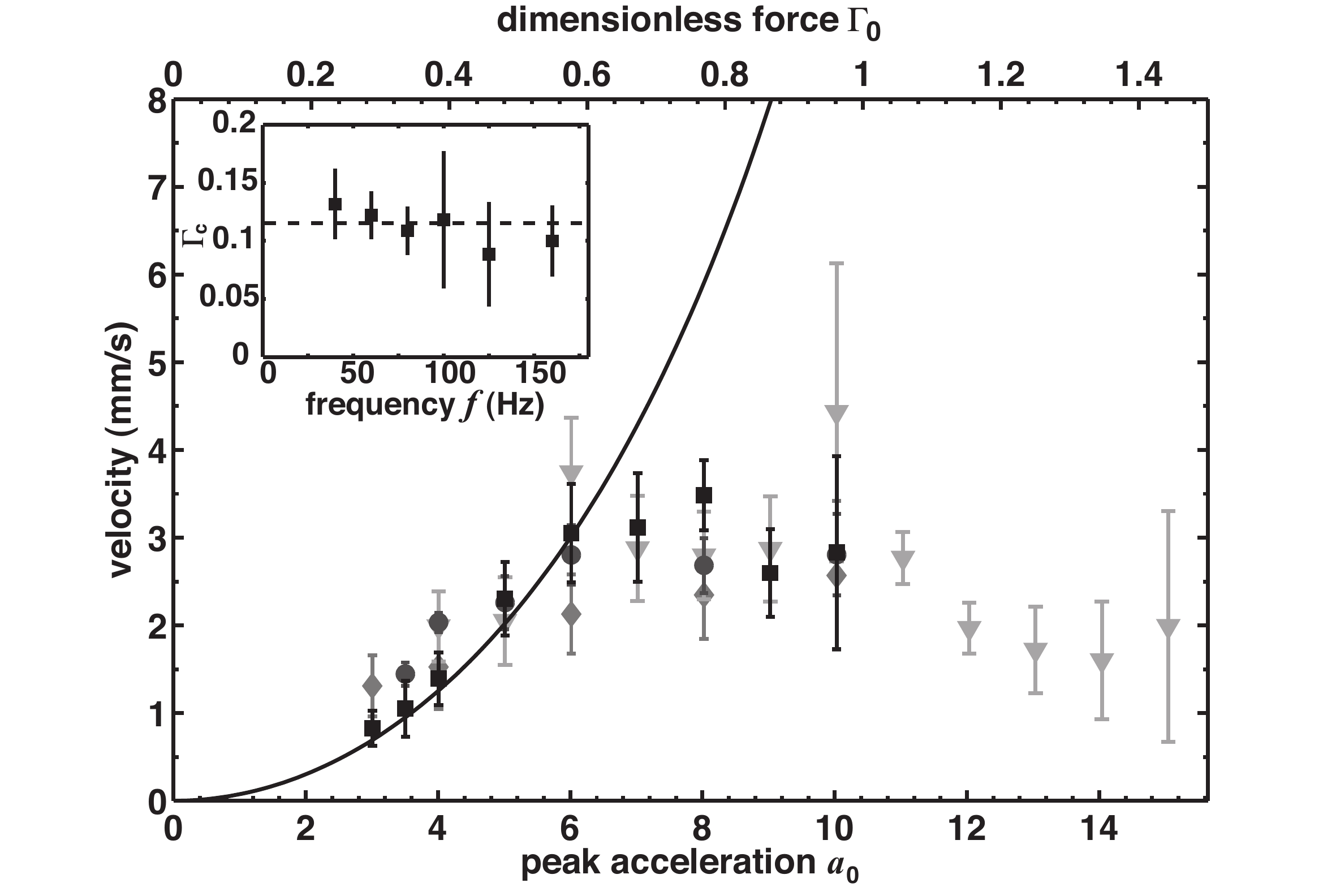}
\caption{{\bf Experimental measurements of the walkers' velocity
 for varying peak substrate acceleration $a_0$ and dimensionless force $\Gamma_0$ (eq. \ref{E_gam})}. These data were obtained for a walker with $R_1 = 0.3$ mm and $R_2 = 0.2$ mm using the following vibration frequencies: 60 Hz (diamonds), 70 Hz (circles), 80 Hz (triangles) and 90 Hz (squares).
 These structures were assembled on a glass slide wetted with Si AK5. The solid line represents a fit of equation \ref{Velocity} to the experimental data where the coefficient of sliding friction, $\kappa = 0.033$ s$/$mm, is the only fit parameter.
{\bf Inset:} the dimensionless force at the onset of motion, $\Gamma_{\mbox{c}}$, as a function of
the vibration frequency $f$. $\Gamma_{\mbox{c}}$ was measured by gradually decreasing the peak plate acceleration $a_0$ and waiting for the walker to stop moving. The dashed line is a guide to the eye situated at the mean value of $\Gamma_{\mbox{c}}$. These experiments were performed on the surface of a glass prism wetted with Si AK5. The error bars represent the range of data resulting from multiple experimental measurements. } \label{Ffig2}
\end{center}
\end{figure}
\noindent
%

%
\begin{figure}[htb!]
\begin{center}
\includegraphics[width=5in]{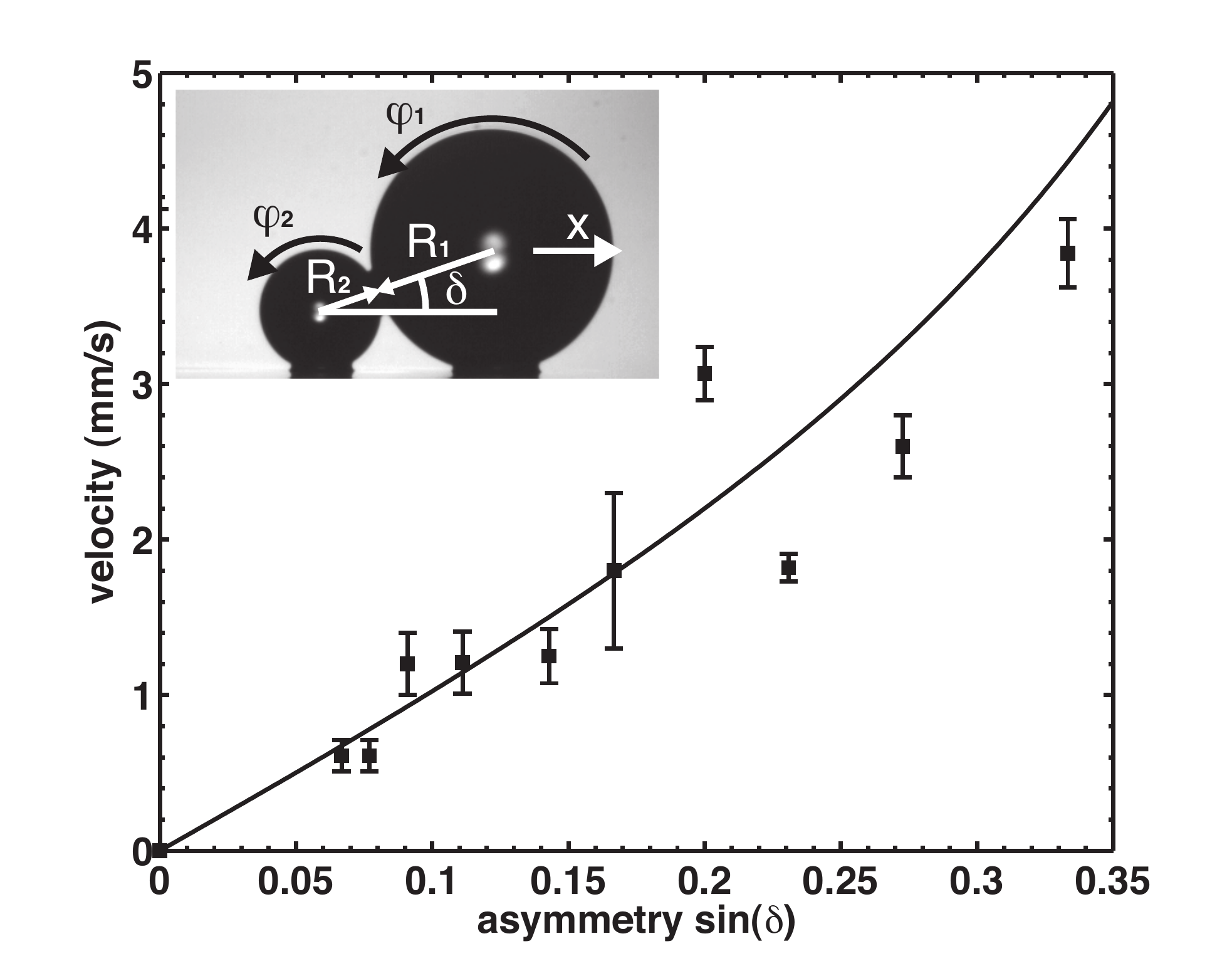}
\caption{{\bf Quantification of the asymmetry and dependence of the
walkers' velocity on the asymmetry.} {\bf Inset:} Schematic
view of a walker composed of two spheres, 1 and 2. We characterize the
walkers' asymmetry in terms of the angle $\delta$ formed
by connecting the centers of the spheres with the horizontal, and our asymmetry parameter 
is defined as $\sin{\delta} = \frac{R_1-R_2}{R_1+R_2}$. The vibration axis of the
substrate is denoted as $x$, and the angular displacements of the
spheres at each shake are denoted as  $\varphi_1$ and $\varphi_2$.
{\bf Main panel:} The walkers' velocity as a function of
the asymmetry parameter $\sin{\delta}$ (squares). These
measurements were performed for walkers constructed from precision
spheres with 0.2, 0.25, 0.3, 0.35 and 0.4 mm radii on the surface of a
glass prism wetted with Si AK5. The vibration
frequency was kept fixed at 80 Hz and the peak acceleration was
varied in order to keep the dimensionless force $\Gamma_0$ (eq.
\ref{E_gam}) fixed at 0.5. The error bars represent the range of data resulting
from multiple experimental measurements. The solid curve shows the expected
theoretical dependence of the walkers' velocity on the asymmetry parameter
$\sin{\delta}$ from equation \ref{Velocity} with $\kappa = 0.033$ s$/$mm 
(the same value as was used in figure \ref{Ffig2}).} \label{Ffig3}
\end{center}
\end{figure}
%

%
\noindent


\begin{thebibliography}{99}

\bibitem{HuangAPL} Huang, T. J. {\it et al.}, A nanomechanical device based on linear molecular motors, {\it Appl. Phys. Lett } {\bf 85,} 5391 (2004).

\bibitem{DreyfusNature} Dreyfus, R. {\it et al.}, Microscopic artificial swimmers, {\it Nature} {\bf 437,} 862 (2005).

\bibitem{BernaNatMat} Bern\'a, J. {\it et al.}, Macroscopic transport by synthetic molecular machines, {\it Nature Materials} {\bf 4,} 704 (2005).

\bibitem{BrowneNatNano} Browne, W. R. \& Feringa B. L., Making molecular machines work, {\it Nature Nanotechnology} {\bf 1,} 25 (2006).

\bibitem{YuPhysFlu} Yu, T. S., Lauga, E. \& Hosoi, A. E., Experimental investigation of elastic tail propulsion at low Reynolds number, {\it Phys. Fluids} {\bf 18,} 091701 (2006).

\bibitem{QianPRL} Qian, B., Powers, T. R. \& Breuer, K. S., Shape transition and propulsive force of an elastic rod rotating in a viscous fluid, {\it Phys. Rev. Lett.} {\bf 100,} 078101 (2008).

\bibitem{TiernoPRL} Tierno, P., Golestanian, R., Pagonabarraga, I. \& Sagu\'es, F., Controlled swimming in confined fluids of magnetically activated colloidal rotors, {\it Phys. Rev. Lett.} {\bf 101,} 218304 (2008).

\bibitem{ToonderLoC} den Toonder, J. {\it et al.}, Artificial cilia for active micro-fluidic mixing, {\it Lab on a Chip} {\bf 8,} 533 (2008).

\bibitem{ZerroukiNature} Zerrouki, D. {\it et al.}, Chiral colloidal clusters, {\it Nature} {\bf 455,} 380 (2008).

\bibitem{GhoshNano} Ghosh, A. \& Fischer P., Controlled propulsion of artificial magnetic nanostructured propellers, {\it Nano Letters} {\bf 9,} 2243 (2009).

\bibitem{ZhangAPL} Zhang, L. {\it et al.}, Artificial bacterial flagella: fabrication and magnetic control, {\it Appl. Phys. Lett } {\bf 94,} 064107 (2009).

\bibitem{AransonSwimmingsnakes} Snezhko, A., Belkin, M., Aranson, I. S. \& Kwok, W.-K., Self-assembled magnetic surface swimmers, {\it Phys. Rev. Lett.} {\bf 102,} 118103 (2009).

\bibitem{WrightDimer} Wright, H. S., Swift, M. R. \& King, P. J., Migration of an asymmetric dimer in oscillatory fluid flow, {\it Phys. Rev. E} {\bf 78,} 036311 (2008).

\bibitem{ChangNatMat} Chang S. T., Paunov, V. N., Petsev, D. N. \& Velev, O. D., Remotely powered self-propelling particles and micropumps based on miniature diodes, {\it Nature Materials} {\bf 6,} 235 (2007).

\bibitem{AransonStaticsnakes} Snezhko, A., Aranson, I. S. \& Kwok, W.-K., Surface wave assisted self-assembly of multidomain magnetic structures, {\it Phys. Rev. Lett.} {\bf 96,} 078701 (2006).


\bibitem{KudrolliDimer} Dorbolo, S., Volfson, D., Tsimring, L. \& Kudrolli, A., Dynamics of a bouncing dimer, {\it Phys. Rev. Lett.} {\bf 95,} 044101 (2005).

\bibitem{KudrolliRod}  Kudrolli, A., Lumay, G., Volfson, D. \& Tsimring, L. S., Swarming and swirling in self-propelled polar granular rods, {\it Phys. Rev. Lett.} {\bf 100,} 058001 (2008).

\bibitem{RosatoPRL} Rosato, A.,Strandburg,  K. J., Prinz,  F. \& Swendsen, R. H., Why the brazil nuts are on top: size segregation of particulate matter by shaking, {\it Phys. Rev. Lett.} {\bf 58,} 1038 (1987).

\bibitem{Fingerle08} Fingerle,  A., Roeller, K., Huang,  K., \& Herminghaus,  S., Phase transitions far from equilibrium in wet granular matter, {\it New J. Phys.} {\bf 10,} 053020 (2008).

\bibitem{McFarlane} McFarlane, J. S., Tabor D., Adhesion of solids and the effect of surface films, {\it Proc. R. Soc. Lond. A} {\bf 202,} 224 (1950).

\bibitem{SuppMat} see Supporting Material, Theory of self-assembled granular walkers.

\bibitem{Olsson} Olsson, H., {\AA}str\"om, K.J., Canudas de Wit, C., G\"afvert, M. \& Lischinsky, P., Friction models and friction compensation, {\it Eur. J. Control} {\bf 4,} 176 (1998).

\bibitem{Ludema} Ludema, K. C., Friction, {\it Modern tribology handbook vol. 1} (CRC Press, edited by B. Bushan, 2001).

\bibitem{Mo09} Mo. Y., Turner, K. T., \& Szlufarska, I. , Friction at the nanoscale, {\it Nature} {\bf 457,} 1116 (2009).

\bibitem{Lachlan} McLachlan, N.W., {\it Theory and Application of Mathieu Functions} (Oxford University Press, London, 1947).

\end{thebibliography}
\end{document}